\documentclass[aps,prl,twocolumn,showpacs]{revtex4}
\usepackage{amsmath,amssymb,latexsym}
\usepackage[english]{babel}
\usepackage{graphicx}

\newtheorem{lemma}{Lemma}
\newtheorem{theorem}{Theorem}
\newcommand{\tachar}[1]{
\setbox4=\hbox{\ }
\setbox3=\hbox{#1}
\hbox{#1}
\kern -\wd3 \kern -\wd4
\raise 0.3\ht3 \hbox{ \vrule width \wd3 height 0.5pt}
}
\newcommand{\tr}{\textrm{Tr}}

\newcommand{\la}{\lambda}
\newcommand{\bea}{\begin{eqnarray}}
\newcommand{\eea}{\end{eqnarray}}
\def\ket#1{|#1\rangle}
\def\bra#1{\left< #1\right|}
\newcommand{\one}{\mbox{$1 \hspace{-1.0mm}  {\bf l}$}}

\begin{document}

\title{Complete set of operational measures for the characterization of $3-$qubit entanglement}
\author{J. I. de Vicente, T. Carle, C. Streitberger and B. Kraus}
\affiliation{Institut f\"ur Theoretische Physik, Universit\"at Innsbruck, Technikerstr.\ 25, 6020 Innsbruck, Austria}

\begin{abstract}
We characterize the entanglement contained in a pure
three--qubit state via operational entanglement measures. To this end we
derive a new decomposition for arbitrary $3$--qubit states which is
characterized by five parameters (up to local unitary operations). We
show that these parameters are uniquely determined by bipartite
entanglement measures. These quantities measure the
entanglement required to generate the state following a particular preparation procedure and have a clear physical meaning.
Moreover, we show that the classification of states obtained in this way
is strongly related to the one obtained when considering general
local operations and classical communication.
\end{abstract}


\maketitle

Entanglement is at the core of many of the applications of quantum information theory and quantum computation and plays a key role in the foundations of quantum mechanics. Therefore, a great amount of theoretical effort has been performed in recent years to grasp this phenomenon, in particular regarding its characterization and quantification as well as its convertibility properties \cite{reviews}. Whereas bipartite entanglement is well understood, multipartite entanglement is much more subtle. In fact, our understanding of the nonlocal properties of many-body states is far from complete even in the simplest case of just three subsystems. Our knowledge of bipartite entanglement stems from the fact that in the asymptotic limit of many copies of any given state, there is a unique optimal rate at which it can be \textit{reversibly} transformed into the maximally entangled state \cite{bipartite}. However, such an approach seems formidable in the multipartite regime \cite{mregs}. Nevertheless, different classes of multipartite entangled states have been identified according to their convertibility properties \cite{ghzw} and several entanglement measures have been proposed like the tangle \cite{ckw}, the localizable entanglement \cite{Popp}, and the entanglement of assistance \cite{ea,ca} to cite a few. A fundamental property of entanglement is that it is invariant under local unitary (LU) operations. This has also led to the study of complete sets of (polynomial) invariants under this kind of operations \cite{LUinvariants} and the necessary and sufficient conditions for LU--equivalence have recently been provided \cite{barbara}. However, a complete classification of LU classes with operationally meaningful measures was still lacking. In this paper we solve this problem for the case of three qubits. That is, we provide a complete set of operational entanglement measures which characterize uniquely all 3--qubit states with the same entanglement properties. These measures, which are easily computable, characterize the different forms of bipartite entanglement involved in the generation of the state following a particular preparation procedure. Our results provide a physical classification of pure multipartite entanglement and, hopefully, will pave the way for new applications of many-body states in the light of quantum information theory.

After removing all the free parameters due to LU operations, two-qubit pure states have one non-local parameter: the Schmidt coefficient. It captures all the information about the entanglement of the state. For $3$--qubit pure states the number of nonlocal parameters is five \cite{LUparameters}. Thus, one would expect five measures to characterize entanglement in this case. Surprisingly, one needs to consider a set of six LU--invariants to characterize the LU--equivalence classes \cite{acin}. This is because of the counterintuitive fact that there exist states which are not LU--equivalent to its complex conjugate \cite{barbara,acin} (which will be taken with respect to the computational basis and will be denoted by $^*$). The sixth invariant is required to discriminate these states. The class of states (not) being LU--equivalent to its complex conjugate will be referred to as CLU (NCLU). The existence of NCLU states is a striking feature of multipartite entanglement which does not exist in the bipartite setting. Since complex conjugation corresponds to the redefinition of the complex unit, $|\psi\rangle$ and $|\psi^*\rangle$ will have the same entanglement properties \cite{barbara}. Thus, one cannot expect operationally meaningful measures to discriminate  between these two states. 
However, as we will show, it is possible to identify states as CLU or NCLU by operational tasks.

The outline of the remainder of the paper is the following.
First, we review some known results and introduce our notation. Then, we derive a decomposition of $3$--qubit states, which is characterized by five parameters. This form leads us naturally to a physical process  generating arbitrary 3--qubit states. Next, we characterize this process by five operational entanglement measures and show that these five measures together with one binary measure characterize the LU--equivalence classes of $3$--qubit states (up to complex conjugation).  Moreover, we show that the CLU--class can be identified by the values these measures take. Last, we show that the characterization of the LU--equivalence classes within CLU is strongly related to the characterization obtained by considering local operations and classical communication (LOCC).

Bipartite entanglement will be measured by the Von Neumann entropy of the reduced state, $E(|\psi\rangle_{12})=S(\rho_1)$, and its convex roof extension for mixed states, the entanglement of formation $E(\rho)=\min_{\{\pi_j,|\phi_j\rangle\}}\sum_j\pi_jE(\phi_j)$ for $\rho=\sum_j\pi_j|\phi_j\rangle\langle\phi_j|$. For two--qubit states, $E=\mathfrak{E}(C)$ is a simple monotonously increasing function of the so called concurrence, which is defined as $C(\psi)=|\langle\tilde{\psi}|\psi\rangle|$, where $|\tilde{\psi}\rangle=\sigma_y\otimes\sigma_y|\psi^*\rangle$ and $\sigma_{x,y,z}$ denote the Pauli operators, and with the convex roof construction for mixed states \cite{concurrence}. An important concept used here is entanglement of assistance. One of the parties (say 1) assists the other two (2 and 3) in obtaining a particular bipartite entangled state by performing measurements on his particle (after possibly adding auxiliary qubits). Let $|\psi\rangle=\sqrt{p} |0\rangle_1|\psi_0\rangle_{23}+\sqrt{1-p} |1\rangle_1|\psi_1\rangle_{23}$ be the Schmidt decomposition of $|\psi\rangle$ in the $1|23$ splitting \cite{Nielsen}. 
The states, $|x_i\rangle$, $23$ obtain with probability $q_i$ after a measurement in $1$ are then given by
\begin{equation}\label{meas}
\sqrt{q_i}|x_i\rangle=\sum_{j=0}^1U_{ij}\sqrt{p_j}|\psi_j\rangle,
\end{equation}
where $U^\dag U=\one$. Here, $i=1,\ldots, d$, where $d$ equals the effective dimension of the subspace in which 1 measures after the (possible) addition of ancillas.
Thus, the states, $|x_i\rangle$, 2 and 3 can get in this way are given by all ensemble decompositions of $\rho_{23}$, $\{\pi_j,|\phi_j\rangle\}$, $\pi_j$ yielding the probability of obtaining outcome $j$ \cite{Nielsen}. Hence, the entanglement of assistance, which is defined as the maximum entanglement on average one can generate by this procedure \cite{ea}, is then given by $E^a_{23}=\max_{\{\pi_j,|\phi_j\rangle\}}\sum_j\pi_jE(\phi_j)$. Let us now introduce the symmetric matrix $\tau$, $\tau_{ij}=\sqrt{p_ip_j}\langle\tilde{\psi_i}|\psi_j\rangle$, and the shorthand notation $c_i=\langle\tilde{\psi_i}|\psi_i\rangle$ and $\tilde{c}=\langle\tilde{\psi_0}|\psi_1\rangle$. Unless stated otherwise we will choose the global phases of $|\psi_0\rangle$ and $|\psi_1\rangle$ such that both $c_0$ and $c_1$ are nonnegative.
There exist closed formulas for both the concurrence $C(\rho_{23})$ and the concurrence of assistance $C^a(\rho_{23})$ in terms of the singular values of $\tau$ \cite{ca,concurrence}. 

We use now the notion of assistance to show that any 3--qubit state can be written as an equal superposition of biseparable states which contain the same amount of entanglement \footnote{In the following we consider only truly tripartite entangled states.}. I.\ e.\ it can be written up to LUs as
\begin{equation}\label{sforma}
|\psi\rangle=\frac{1}{\sqrt{2}}(|0\rangle_1|\psi_s\rangle_{23}+|1\rangle_1U_2\otimes U_3|\psi_s\rangle_{23}).
\end{equation}
Here, $|\psi_s\rangle=a\ket{00}+b\ket{11}$ with $a\geq b$ has Schmidt decomposition and $U_2=Z(\alpha)Y(\beta)Z(\gamma)$ and $U_3=Y(\beta')$ with $Y,Z(\xi)=e^{i\xi\sigma_{y,z}}$ respectively. Thus, any state is characterized by the five parameters, $\{E(\psi_s), \alpha, \beta, \gamma, \beta'\}$, where $E(\psi_s)\in [\mathfrak{E}(C_{23}),\mathfrak{E}(C^a_{23})]$, $\alpha,\gamma\in(-\pi/2,\pi/2]$ and $\beta,\beta'\in[0,\pi/2]$ (see Appendix A). 

To see that any state can be written as in Eq.\ (\ref{sforma}) (up to LUs), we consider a measurement on qubit $1$ such that both outcomes are equally likely, i.e. $q_1=q_2=1/2$ and $E(\ket{x_1})=E(\ket{x_2})$ (see Eq.\ (\ref{meas})) \footnote{This can be seen to be equivalent to measurements on a maximally entangled basis with equiprobable outcomes when an ancillary qubit is given to 1.}. Since $|\psi_0\rangle\bot|\psi_1\rangle$ we find that the unitary in Eq.\ (\ref{meas}) can be taken without loss of generality of the form $U=\frac{1}{\sqrt{2}} \left(
\begin{array}{cc}
1 & \exp(i\omega) \\
-\exp(-i\omega) & 1 \\
\end{array}
\right)$. Then $C(|x_{0,1}\rangle)=|pc_0+(1-p)c_1e^{2i\omega}\pm2\sqrt{p(1-p)}\tilde{c}e^{i\omega}|$ and, thus, $E(|x_0\rangle)=E(|x_1\rangle)$ iff
\begin{equation}\label{phialpha}
\omega=\arctan\left(\frac{pc_0+(1-p)c_1}{pc_0-(1-p)c_1}\cot\arg \tilde{c}\right).
\end{equation}
Since this equation has a solution for any given value of $\arg\tilde{c}$, this proves the statement. The particular form of the unitaries, $U_2,U_3$, is achieved by taking the Euler ZYZ decomposition and using the fact that $Z(\xi)\otimes \one|\psi_s\rangle=\one\otimes Z(\xi)|\psi_s\rangle$. 
Notice that the decomposition (\ref{sforma}) (i.\ e.\ the choice of $\{E(\psi_s), \alpha, \beta, \gamma, \beta'\}$) can be made unique (see Appendix A).

The decomposition (\ref{sforma}) shows that any 3--qubit state can be generated in the following way (see Fig.\ $1$): Qubit $1$ is initialized in the state $|+\rangle$ and qubits $2$ and $3$ are prepared in the state $\ket{\psi_s}$ \footnote{$\ket{\psi_s}$ can be generated via a controlled gate, however, this would not change any of the following argumentation.}. Then qubit $1$ is attached to qubit $2$ via the controlled nonlocal unitary $U_c^{12}$ and to qubit $3$ via $U_c^{13}$ where $U_c^{1i}=|0\rangle_1\langle0|\otimes\one_i+|1\rangle_1\langle1|\otimes U_i$, i.\ e.\ $|\psi\rangle=U_c^{13}U_c^{12}|+\rangle_1|\psi_s\rangle_{23}$. Note that $[U_c^{12},U_c^{13}]=0$.

\begin{figure}[h!]
\begin{center}
  \includegraphics[height=0.13\textheight]{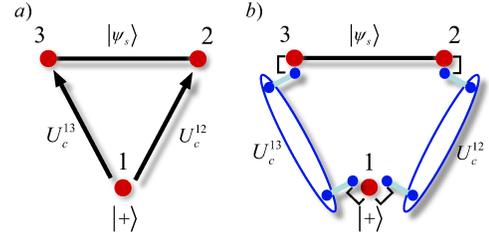}
  \caption{a) The decomposition (\ref{sform}) of an arbitrary $3$--qubit state: $|\psi\rangle=U_c^{13}U_c^{12}|+\rangle_1|\psi_s\rangle_{23}$ and b) the generation process: each party makes a Bell measurement on its qubit and on one of its share of the corresponding CJ state}.\label{fig1}
\end{center}
\end{figure}

Let us now briefly discuss a method to generate the state using entanglement. It is a well--known fact that any nonlocal map, $\mathcal{E}$, corresponding to a state $\ket{\Psi_{\cal E}}$ via the Choi–
Jamiolkowski (CJ) isomorphism  can be (probabilistically) implemented using
a system prepared in the state $\ket{\Psi_{\cal E}}$ and local Bell--measurements \cite{entanglingops} (see Fig.\ $1$). Since the implementation of each control gate, $U_c^{12}$ and $U_c^{13}$, requires four local Bell--measurements, we obtain, depending on the measurement result, one out of the $4^4$ different states, $\{U_c^{13}\sigma_{i_1}^1 \otimes \sigma_{i_2}^3 U_c^{12}  \sigma_{i_3}^1\otimes \sigma_{i_4}^2|+\rangle_1|\psi_s\rangle_{23}\}$ where $\sigma_0=\one$. Due to the symmetry of these states, it can be shown that this set coincides up to LUs to the set $\mathcal{S}_\psi=\{U_c^{13}U_c^{12}\sigma_i^2|+\rangle_1|\psi_s\rangle_{23}\}$, containing only four states (see Appendix B).  Each state in this set occurs with equal probability.
Note that if the aim was to generate any other state within $\mathcal{S}_\psi$, one would again, depending on the measurement outcome, end up with one of the states in $\mathcal{S}_\psi$. That is, the set $\mathcal{S}_\psi$ is closed under this generation process
. In the following we derive bipartite entanglement measures, which characterize such a generation process completely. This will finally lead to our main result: the operational characterization of the entanglement contained in 3-qubit states.

In order to characterize this process we need on the one hand $E(\ket{\psi_s})$ and on the other we need to characterize the entanglement properties of 2-qubit controlled unitaries. Recall that any 2-qubit unitary can be written as $V_1\otimes V_2U_{nl}V_3\otimes V_4$ \cite{nlunits}. Here, the nonlocal part $U_{nl}$ is diagonal in the Bell basis and depends in general on three parameters. Let us consider the control gate $U_c^{12}$ and write without loss of generality $U_2=WZ(2\gamma)W^\dag$, where $W=Z(\tilde{a})Y(\tilde{b})$ with $\tilde{a},\tilde{b}\in\mathbb{R}$. Then, in this case it can be easily checked that the nonlocal content $U_{nl}$ of $U_c^{12}$ depends on a single parameter
, $\gamma$, which is in turn uniquely characterized by $E_{imp}(U_c^{12})$, which denotes the amount of entanglement (in the splitting $1|2$) contained in the CJ--state corresponding to $U_c^{12}$, i.\ e.\ the amount of entanglement needed to \textit{implement} $U_c^{12}$. Thus, knowing just $U_{nl}$ is not enough to characterize $U_c^{12}$ as this does not specify the other parameters $\tilde{a}$ and $\tilde{b}$, which depend on the LUs $\{V_i\}$. These are as well relevant to characterize the entanglement properties of $U_c^{12}$ as can be easily understood as follows. 
Let $U_c^{1\rightarrow2}=|0\rangle\langle0|\otimes\one+|1\rangle\langle1|\otimes U_2$ and $U_c^{1\leftarrow 2}=\one\otimes|0\rangle\langle0|+U_2\otimes|1\rangle\langle1|$ (i.\ e.\ the same controlled unitary but with different control qubit). It can be seen that the nonlocal parts of these operations are equivalent. Thus, the choice of the local unitaries $V_i$ determines which qubit is the control qubit: $U_c^{1\rightarrow 2}=\one \otimes WZ(\gamma)U_{nl} \one\otimes W^\dag$ and $U_c^{1\leftarrow 2}=WZ(\gamma) \otimes \one U_{nl} W^\dag \otimes \one$. Obviously the information about the control qubit is important, e.g. the CNOT gate acting on $|+0\rangle$, generates either a product state or maximally entangled state depending on which qubit is the control qubit. In summary, $E_{imp}$ alone does not characterize the entanglement properties of a controlled unitary. One needs further measures to characterize the local unitaries $V_i$. This will allow to specify $U_c^{12}$ uniquely. As we will see, this will make it also possible to characterize our generation process and will lead to a complete classification of 3--qubit entanglement. The most natural choice is then the entanglement created after the implementation of the controlled unitary \cite{LOCCunitaries}. Since, in the considered case, the controlling direction is not specified by $U_{nl}$ a reasonable choice is the entanglement gained with the two possibilities, $E_{gain}^\rightarrow(U_c^{12})$ and $E_{gain}^\leftarrow(U_c^{12})$, for some physically motivated choice of input state. It is easy to see that these measures will specify $\tilde{a}$ and $\tilde{b}$.

Now we are in the position to state our main result, the operational characterization of the LU classes up to complex conjugation of 3-qubit states. First, we have that the set $\mathcal{S}_\psi$, that is the generation process described above, is uniquely characterized by the values of the following five bipartite operational measures \footnote{Since the CJ state corresponding to $U_c^{1i}$ is $(|00\rangle_1|\phi^+\rangle_i+|11\rangle_1 U_i\otimes\one|\phi^+\rangle_i)/\sqrt{2}$, we have $E_{imp}(U_c^{1i})=S((|\phi^+\rangle\langle\phi^+|+U_i\otimes\one |\phi^+\rangle\langle\phi^+|U_i^\dag\otimes\one)/2)$, $E_{gain}^\rightarrow(U_c^{12})=E_{1|23}(U_c^{12}|+\rangle_1|\psi_s\rangle_{23})=E_{1|23}(1/\sqrt{2}(|0\rangle_1|\psi_s\rangle_{23}+|1\rangle_1U_2\otimes \one|\psi_s\rangle_{23}))$, and $E_{gain}^\leftarrow(U_c^{12})=E_{1|23}(U_c^{21}|+\rangle_1|\psi_s\rangle_{23})=E_{1|23}(a|+\rangle_1|00\rangle_{23}+bU_2(|1\rangle_1)|11\rangle_{23}))$. Notice that $U_c^{13}$ is characterized by just one parameter, e.g. $E_{imp}(U_c^{13})$.}
\begin{align}\label{measures}
&E_1=E(\psi_s),\quad E_2=E_{imp}(U_c^{12}),\quad E_3=E_{imp}(U_c^{13}),\nonumber\\
&E_4=E_{gain}^\rightarrow(U_c^{12}),\quad E_5=E_{gain}^\leftarrow(U_c^{12}),
\end{align}
where $E_{gain}^\rightarrow(U_c^{12})$ ($E_{gain}^\leftarrow(U_c^{12})$) measures the amount of entanglement generated between system 1 and systems 2 and 3 by applying the control gate $U_c^{12}$ to the input state $|+\rangle_1|\psi_s\rangle_{23}$ and the direction of the arrow indicates whether system 1 ($\rightarrow$) or system 2 ($\leftarrow$) is the control qubit. To see that this is indeed the case, we need to verify that $\{E(|\psi_s\rangle), \alpha, \beta, \gamma, \beta'\}$ in Eq.\ (\ref{sforma}) can be obtained from the values of $\{E_i\}$. 
Since for all the bipartite entanglement measures at least one subsystem belongs to an effective two-dimensional space, they are in one-to-one correspondence with the concurrence, and therefore with the trace of the squared reduced density matrix. Thus, $E_3$ is characterized by the value of $(1+|\langle\phi^+|Y(\beta')\otimes\one|\phi^+\rangle|^2)/2=(1+\cos^2\beta')/2$, which shows the one-to-one correspondence of $E_3$ and $\beta'$ (recall that $\beta'\in[0,\pi/2]$). Analogously, $E_2$, $E_4$ and $E_5$ are respectively in unique correspondence with $\cos^2\beta\cos^2(\alpha+\gamma)$, $\cos^2\beta[(a^2-b^2)^2+4a^2b^2\cos^2(\alpha+\gamma)]$ and $a^4+b^4+2a^2b^2[\cos^2(\alpha+\gamma)\cos^2\beta+\sin^2(\alpha-\gamma)\sin^2\beta]$. 
Clearly then, $E_2$ and $E_4$ determine $\beta$ uniquely and $\alpha+\gamma$ modulo $\pi$, which has only (at most) four LU inequivalent solutions which are compatible with $E_5$: $\{\alpha,\gamma\}$, $\{\gamma,\alpha\}$, $\{\alpha\pm\pi/2,\gamma\pm\pi/2\}$ and $\{\gamma\pm\pi/2,\alpha\pm\pi/2\}$, where the signs in the last two cases are chosen such that the phases are in the desired region. This set of states coincides with $\mathcal{S}_\psi$ \footnote{Note that for $E_1=1$, $E_2=E_4$ and, because of our choice of the decomposition (2) $E_3=0$ (see Appendix A). As in Appendix A, changing one of the phases by $\pm \pi$ leads to an LU--equivalent state, since $Z(\pi)=Y(\pi)=-\one$. Let us define $\ket{\psi(\alpha,\beta)}=U_{c}^{12}(\alpha,\beta)\ket{+}\ket{\psi_s}$. Note that changing the sign of $\alpha$ or $\beta$ correspond to LU--equivalent states. Then, it is easy to see that the only states (up to LU) which are compatible with $E_2$ and $E_5$ are $\ket{\psi( \alpha, \beta)}$ and $\ket{\psi(\beta,\alpha)}$.  These states can be shown to be LU--equivalent using similar methods. Thus, in this case $\mathcal{S}_\psi=\{\ket{\psi}\}.$}. 
Hence, a one-to-one correspondence is made between the entanglement measures $\{E_i\}$ and the set of states obtained through our generation process.

Let us now have a closer look at the set
$\mathcal{S}_\psi=\{U_c^{13}U_c^{12}\sigma_i^2|+\rangle_1|\psi_s\rangle_{23}\}$. Note that $\mathcal{S}_\psi=\{\ket{\psi},\ket{\psi^\ast},\ket{\psi'},\ket{\psi'^\ast}\}$, where $\ket{\psi'}=U_c^{13}U_c^{12}\sigma_z^2|+\rangle_1|\psi_s\rangle_{23}$.
As discussed in the introduction there is no operational measure to distinguish a state from its complex conjugate, hence it is not surprising to have this degeneracy. On the contrary, one could ask if there is an operational procedure to further distinguish $|\psi\rangle$ and $|\psi'\rangle$. This extra bit of information can be easily obtained by considering the value of the final entanglement generated in the splitting $1|23$, $E_{1|23}(\psi)$. It can be seen that $E_{1|23}(\psi)=E_{1|23}(\psi')$ iff $|\psi\rangle$ or $|\psi^*\rangle$ are LU--equivalent to $|\psi'\rangle$ (see Appendix C). Thus, define the binary measure $E_6\equiv E_6^{S_\psi}$ as $E_6=0$ ($E_6=1$) when $E_{1|23}$ takes the minimal (maximal) possible value inside $\mathcal{S}_\psi$. The operational measures $\{E_i\}_{i=1}^6$ characterize uniquely (up to complex conjugation) the LU classes of 3-qubit states, and, therefore, their entanglement properties.

Despite the fact that $E_i(\psi)=E_i(\psi^*)$ $\forall i$, as expected, our approach can, nevertheless, provide also information on this issue: the values the measures $\{E_i\}$ take depend on whether $|\psi\rangle$ belongs to the CLU or to the NCLU class. More explicitly, we have that $|\psi\rangle\in$ CLU iff either $E_1=\mathfrak{E}(C_{23})$ or $E_1=\mathfrak{E}(C_{23}^a)$ (while $\mathfrak{E}(C_{23})<E_1<\mathfrak{E}(C_{23}^a)$ for NCLU states) (see Appendix D). In other words, it takes the same bipartite entanglement resources to generate a state and its complex conjugate, but while for CLU states the isolated qubit is attached to some bipartite state whose entanglement is either the maximal or minimal possible, the NCLU states require some intermediate amount of entanglement.

Furthermore, a refined classification of the CLU class can be achieved by considering the value of $E_1$. We have the following subclasses of states $|\psi\rangle$ with $E_1=E_1(\psi)$ (in what follows $c\in\mathbb{R}$ and $\phi_i\in\mathbb{R}^2$) (see Appendix E):
\begin{itemize}
\item[1.] $E_1=\mathfrak{E}(C_{23})=\mathfrak{E}(C_{23}^a)$ iff $|\psi\rangle\in$ W--class \cite{ghzw}.
\item[2.] $E_1=\mathfrak{E}(C_{23}^a)\neq\mathfrak{E}(C_{23})$ iff $|\psi\rangle\propto
|000\rangle+c|\phi_1 \phi_2 \phi_3\rangle$.
\item[3.] $E_1=\mathfrak{E}(C_{23})\neq\mathfrak{E}(C_{23}^a)$ iff $|\psi\rangle\propto|000\rangle+e^{ic}|\phi_1 \phi_2 \phi_3\rangle$.
\item[4.] $\exists|\psi_s\rangle,|\psi_s'\rangle$ for Eq.\ (\ref{sform}) such that $E_1(\ket{\psi_s})=\mathfrak{E}(C_{23})$ $\land$ $E_1(\ket{\psi^\prime_s})=\mathfrak{E}(C_{23}^a)$ ($\mathfrak{E}(C_{23})\neq \mathfrak{E}(C_{23}^a)$) iff $|\psi\rangle\propto(|000\rangle+|\phi_1 \phi_2 \phi_3\rangle)$.
\end{itemize}
This can be regarded as a classification beyond the stochastic LOCC paradigm \cite{ghzw} since the last three classes correspond to GHZ-type states. This classification is related to the classification of states according to deterministic LOCC operations. It follows from the results of \cite{LOCC} that classes 1, 2 and 3 are closed under these transformations, i.\ e.\ deterministic LOCC can only take a state to another state within the same class. Class 4 is composed of those states for which $|\psi_s\rangle$ in the decomposition given by Eq.\ (\ref{sforma}) is not unique. In fact, for these states $E_1$ can take any value between $\mathfrak{E}(C_{23})$ and $\mathfrak{E}(C_{23}^a)$ (cf.\ Appendix E). This richness is directly connected to the LOCC properties of this class of states: these are the only ones that might be transformed to other classes \cite{LOCC} (including NCLU states). Note that, due to the symmetry of the states in any class, this classification is invariant under the considered splitting, that is, if e.g. $E_1=\mathfrak{E}(C_{23})$ in splitting $1\mid 23$, it is so in any other splitting.

In conclusion, we derived a decomposition for 3-qubit pure states, which, as the Schmidt decomposition for bipartite states, can be easily computed. This decomposition leads naturally to a generation process, which is characterized by the bipartite entanglement measures, $\{E_i\}_{i=1}^5$ (Eq.\ (\ref{measures})). The set $\{E_i\}_{i=1}^5$ together with the binary measure $E_6$ forms a complete set of operational measures identifying the different LU--equivalence classes (up to complex conjugation). Hence, the nonlocal properties of $3$--qubits states have been operationally characterized. Even though the measures are bipartite, it should be stressed that this is by no means an account of entanglement across different bipartite splittings.
The maximally entangled state, with $E_i=1$, $\forall i$ is the GHZ--state \footnote{In fact, as explained before, if $E(\ket{\psi_s})=1$, we only need to consider two additional measures to identify the state.}. 
We have also analyzed some features of the classification induced by this set of measures and we have shown that the value of $E_1$ determines whether a state is LU--equivalent to its complex conjugate or not. If this is the case, then four classes can be identified which are related to how the states may transform under deterministic LOCC operations, showing further the physicality of our approach. It is worth remarking that such a connection can already be established from the value of a single measure. It will be interesting to study if the other measures can provide a finer classification of states under these transformations. Although the values of $\{E_i\}$ depend on the chosen initial partition, it is worth remarking that this classification is not (neither of course that of the states with the same entanglement). This reflects the fact that we are dealing with multipartite states and one might choose the partition depending on the particular task one wants to accomplish. Given the operational character of our approach, in the future we will study how to generalize it to more qubits and higher dimensions with the aim to understand the most relevant measures and identify possible new applications of multipartite entangled states. 

The research was funded by the Austrian Science Fund (FWF): Y535-N16 and
F40-FoQus F4011-N16.

\section{Appendix}

\subsection{Appendix A: Mathematical properties of the 3--qubit state decomposition}

In the main text we have proven that any 3--qubit state can be written up to LUs as
\begin{equation}\label{sformapp}
|\psi\rangle=\frac{1}{\sqrt{2}}(|0\rangle_1|\psi_s\rangle_{23}+|1\rangle_1U_2\otimes U_3|\psi_s\rangle_{23}),
\end{equation}
where $|\psi_s\rangle=a\ket{00}+b\ket{11}$ with $a\geq b$ has Schmidt decomposition and $U_2=Z(\alpha)Y(\beta)Z(\gamma)$ and $U_3=Y(\beta')$ with $Y,Z(\xi)=e^{i\xi\sigma_{y,z}}$ respectively. Thus, any state is characterized by the five parameters, $\{E(\psi_s), \alpha, \beta, \gamma, \beta'\}$ \footnote{Notice that $E(\psi_s)$ identifies $|\psi_s\rangle$ uniquely as it has Schmidt decomposition.}. In this section we prove that they can be taken such that $E(\psi_s)\in [\mathfrak{E}(C_{23}),\mathfrak{E}(C^a_{23})]$, $\alpha,\gamma\in(-\pi/2,\pi/2]$ and $\beta,\beta'\in[0,\pi/2]$ and show how the decomposition can be made unique.

We start by proving that $\mathfrak{E}(C_{23})\leq E(\psi_s)\leq\mathfrak{E}(C^a_{23})$. As introduced in the main text it is useful to regard Eq.\ (\ref{sformapp}) as providing a \textit{deterministic} assistance protocol in which party 1 by performing some measurement will induce some entanglement between 2 and 3. More precisely, carrying out a Von Neumann measurement on the basis $\{|0\rangle,|1\rangle\}$ will surely produce 23 to hold a state with entanglement $E(\psi_s)$. The result then follows by taking into account that the concurrence and the concurrence of assistance, which are the extremal values (as measured by the concurrence) that can be obtained by \textit{any} assistance protocol, can be obtained by some deterministic protocol \cite{concurrence,detass}. It is not clear, however, that the particular protocol corresponding to decomposition (\ref{sformapp}) allows certain states to attain any of these extremal values but this turns out to be the case (cf.\ Theorem 1 in Sec.\ IV below).

To see that $\{\alpha, \beta, \gamma, \beta'\}$ can be taken to lie in the regions stated above, first notice that it suffices to consider the first and fourth quadrant since rotations around $\pi$ correspond to the LU $\sigma_z^1$. The rest follows by noticing that the following transformations on $\{\alpha, \beta, \gamma, \beta'\}$ correspond to LUs: $\{\alpha, -\beta, \gamma, -\beta'\}$ (multiplying by $\one\otimes\sigma_z\otimes\sigma_z$), $\{\alpha\pm\pi/2, -\beta, \gamma\pm\pi/2, \beta'\}$ (because $Y(\beta)=-Z(\pi/2)Y(-\beta)Z(\pi/2)$) and $\{-\gamma, -\beta, -\alpha, -\beta'\}$ (multiplying by $\sigma_x\otimes U_2^\dag\otimes U_3^\dag$). Notice that this implies that the transformation to $\{\gamma, \beta, \alpha, \beta'\}$ (as well as $\{-\alpha, \beta, -\gamma, \beta'\}$) corresponds to complex conjugation, which is used in the main text.

The question of uniqueness of the decomposition (\ref{sformapp}) is not crucial for our purposes as one can always define a particular choice of $|\psi_s\rangle$, $U_2$ and $U_3$ to make our measures $\{E_i\}$ well defined. As we have shown in the main text this is enough to characterize operationally the classes of 3--qubit states with the same entanglement. Nevertheless, let us discuss this issue here for the sake of clarity and completeness. First, one may wonder if the choice of $|\psi_s\rangle$ is unique (i.\ e.\ whether there is one single value such that $E(|x_0\rangle)=E(|x_1\rangle)$). This can be easily verified in almost all cases as one can read from Eq.\ (3) in the main text that $\omega$ is uniquely defined (up to the irrelevant addition of $n\pi$ with $n\in\mathbb{Z}$) except when i) $c_0=c_1=0$, ii) $pc_0=(1-p)c_1\neq0$ and $\arg\tilde{c}=\pi/2$ and iii) $\tilde{c}=0$ in which it can take any value. Nevertheless, in case i) $|\psi_s\rangle$ is also unique since $C(|x_{0,1}\rangle)$ does not depend on $\omega$ (in what follows in this case we will choose initially $\omega=0$ so that the the unitaries $U_2$ and $U_3$ can be made unique). On the other hand, in cases ii) and iii) $C(|x_{0,1}\rangle)$ changes with $\omega$ and, hence, different choices for $|\psi_s\rangle$ are possible (which would lead to different values for our measures as $E_1=E(\psi_s)$). This is, for instance, the case of class 4 inside the CLU class and, hence, it seems that there could be some physical meaning behind this fact. Nevertheless, the choice of $|\psi_s\rangle$ can be fixed by taking the one for which $E(\psi_s)$ is maximal. Once the uniqueness of $|\psi_s\rangle$ has been settled one could ask whether then $U_2$ and $U_3$ are unique (with the constraint that they must be respectively of the ZYZ and the Y form). In other words, whether there exist LUs $\{V_i\}$ such that
\begin{align}
V_1&\otimes V_2\otimes V_3\frac{1}{\sqrt{2}}(|0\rangle_1|\psi_s\rangle_{23}+|1\rangle_1U_2\otimes U_3|\psi_s\rangle_{23})\nonumber\\ &=\frac{1}{\sqrt{2}}(|0\rangle_1|\psi_s\rangle_{23}+|1\rangle_1U'_2\otimes U'_3|\psi_s\rangle_{23}).
\end{align}
It can be seen that this is possible iff $V_1\in\{\sigma_i\}_{i=0}^4$ up to a global phase \footnote{Arguing as in the proof of the decomposition (2) in the main text, $V_1$ can be interpreted in the light of an assistance protocol. Thus, it can be readily checked that, in order to preserve the probabilities and the amount of entanglement for the new outcomes, $V_1$ must be either diagonal or off-diagonal (save in the case i) above for which this is fixed by imposing $\omega=0$). Then, it can be seen that in both cases the non-vanishing entries must have either the same or opposite phase so that $U'_2$ and $U'_3$ have unit determinant (i.\ e.\ they can have the Euler ZYZ decomposition).}. Hence, $V_2$ and $V_3$ must be such that either $V_2\otimes V_3|\psi_s\rangle_{23}=|\psi_s\rangle_{23}$ and $V_2\otimes V_3U_2\otimes U_3|\psi_s\rangle_{23}=U'_2\otimes U'_3|\psi_s\rangle_{23}$ or $V_2\otimes V_3|\psi_s\rangle_{23}=U'_2\otimes U'_3|\psi_s\rangle_{23}$ and $V_2\otimes V_3U_2\otimes U_3|\psi_s\rangle_{23}=|\psi_s\rangle_{23}$ (up to some possible change of signs). Then, if $|\psi_s\rangle$ is not the maximally entangled state, since $Z(\xi)\otimes Z(-\xi)$ for any $\xi$ is the most general LU operation that leaves $|\psi_s\rangle$ invariant, it follows that the above considered transformations on $\{\alpha, \beta, \gamma, \beta'\}$ are the \textit{only} ones that can occur. Hence, taking into account that the parameters are fixed to $\alpha,\gamma\in(-\pi/2,\pi/2]$ and $\beta,\beta'\in[0,\pi/2]$ and imposing, for instance, $|\alpha|\geq|\gamma|$ the decomposition is made unique. Notice, however that this is irrelevant as our measures $\{E_i\}$ are invariant under these transformations. On the other hand, if $|\psi_s\rangle$ is a maximally entangled state, then operations of the form $U\otimes U^*$ for any unitary $U$ are in this case the most general LU transformations that leave $|\psi_s\rangle$ invariant. Then, in this case in order to make the decomposition unique and our measures $\{E_i\}$ uniquely defined we take without loss of generality $U_2=Z(\alpha)Y(\beta)$ and $U_3=\one$.

\subsection{Appendix B: Implementation of the controlled unitary operations to generate $\mathcal{S}_\psi$}

As explained in the main text, to apply the non--local gate $U_c^{12}=|0\rangle_1\langle0|\otimes\one_2+|1\rangle_1\langle1|\otimes U_2$ on some state $|\phi\rangle_{12}$, 1 and $2$ share in addition to their system qubits the CJ--state $|\Psi\rangle_{1_a1_b2_a2_b}=(|00\rangle_1|\phi^+\rangle_2+|11\rangle_1 U_2\otimes\one|\phi^+\rangle_2)/\sqrt{2}$ \cite{entanglingops} (see Fig.\ $1$ in the main text). Depending on the measurement outcome they obtain one of the four states $U_c^{12}\sigma_i\otimes\sigma_j|\phi\rangle$ ($\sigma_0$ corresponding to the outcome $\phi^+$, $\sigma_x$ to $\psi^+$, $\sigma_y$ to $\psi^-$ and $\sigma_z$ to $\phi^-$). Thus, implementing $U_{c}^{12}$ and $U_{c}^{13}$ in this way leads to one of the $4^4$ states in the set $\{U_c^{13}\sigma_i^1\otimes\sigma_j^3U_c^{12}\sigma_k^1\otimes\sigma_l^2|+\rangle_1|\psi_s\rangle_{23}\}$.
We show here that, due to the symmetry of the states, all of them are LU--equivalent to one of the four states in $\mathcal{S}_\psi=\{U_c^{13}U_c^{12}\sigma_n^2|+\rangle_1|\psi_s\rangle_{23}\}_{n=0}^4$.
We have
\begin{align}
U_c^{13}\sigma_i^1\otimes\sigma_j^3&U_c^{12}\sigma_k^1\otimes\sigma_l^2|+\rangle_1|\psi_s\rangle_{23}\nonumber\\
&\simeq_{LU}U_c^{13}U_c^{12}\sigma_l^2\otimes\sigma_m^3|+\rangle_1|\psi_s\rangle_{23}\nonumber\\
&\simeq_{LU}U_c^{13}U_c^{12}\sigma_n^2|+\rangle_1|\psi_s\rangle_{23},\label{impl}
\end{align}
for some $m$ and $n$ $\in\{0,\ldots, 3\}$. The reason for this is that commuting the Pauli operations with $U_c^{1i}$ will at most change the sign of both, $\alpha$ and $\gamma$, and/or of $\beta$ and/or of $\beta'$. More precisely, the second line follows from the fact that $\sigma_i^1\ket{+}\in\{\ket{+},\ket{-}\}$ and the fact that $U_c^{13}\sigma_j^1$ is up to LU (from left) either $U_c^{13}$ or $(U_c^{13})^\dagger$, which is up to LU (from left) equivalent to $U_c^{13}\sigma^3_1$. Note that the second case corresponds to changing the sign of $\beta'$. The third line follows from the fact that the change of the sign of $\beta'$ can be accounted for by a change of the sign of $\beta$. Therefore, all output states are LU--equivalent to one of the states presented in Eq.\ (\ref{impl}). Now, since $U_c^{13}U_c^{12}\sigma_n^2|+\rangle_1|\psi_s\rangle_{23}$ equals
$|\psi\rangle$ for $n=0$, $|\psi'^*\rangle$ for $n=1$, $|\psi^*\rangle$ for $n=2$, and $|\psi'\rangle$ for $n=z$, this set coincides with the set $\mathcal{S}_\psi$.

Note that $\mathcal{S}_\psi$ contains \textit{at most} four LU--inequivalent states, since under certain conditions some of these states are LU--equivalent (see main text). 
It can be easily verified that the four possibilities are obtained with equal probability as for the implementation of $U_c^{12}$
\begin{equation}
||_{1_b1_c}\langle\phi^+|_{2_b2_c}\langle\phi^+|\sigma_i^{1_b}\sigma_j^{1_c}\sigma_k^{2_b}\sigma_l^{2_c}|\Psi\rangle_{1_a1_b2_a2_b}|+\rangle_{1_c}|\psi_0\rangle_{2_c3}||^2
\end{equation}
always equals $1/16$ and, hence, does not depend on $i,j,k,l$ and similarly for the posterior application of $U_c^{13}$.

\subsection{Appendix C: Identifying $|\psi\rangle$ and $|\psi'\rangle$ with $E_6$}

We show here in detail that the binary measure $E_6$ allows us to distinguish the two states $|\psi\rangle$ and $|\psi'\rangle$, if they are not LU--equivalent. Recall that $E_6(\ket{\Phi})$ is defined as follows: $E_6(\ket{\Phi})=0$ if $E_{1|23}(\ket{\Phi})=\mbox{min}_{\ket{\Phi_i}\in S_{\psi}}E_{1|23}(\ket{\Phi_i})$ and $E_6(\ket{\Phi})=1$ otherwise. Since $E_{1|23}(\ket{\Phi})=E_{1|23}(\ket{\Phi^\ast})$, for any state $\ket{\Phi}$, the set $\{E_{1|23}(\ket{\Phi_i}), \ket{\Phi_i}\in S_{\psi}\}$ contains at most two values. Using the same reasoning as in the main text, $E_{1|23}$ is uniquely given by the value of
\begin{align}\label{e6}
&|\langle\psi_s|Z(\alpha)Y(\beta)Z(\gamma)\otimes Y(\beta')|\psi_s\rangle|^2=4a^2b^2\sin^2\beta\sin^2\beta'\nonumber\\&\cos^2(\alpha-\gamma)+\cos^2\beta\cos^2\beta'[(a^2-b^2)^2+4a^2b^2\cos^2(\alpha+\gamma)]\nonumber\\
&+4ab\cos\beta\cos\beta'\sin\beta\sin\beta'\cos(\alpha+\gamma)\cos(\alpha-\gamma).
\end{align}

Recall that if $\ket{\psi}$ corresponds to the parameters $(E(\ket{\psi_s}),\alpha,\beta,\gamma,\beta')$ (see Eq.\ (2) in the main text), then $\ket{\psi'}$ corresponds to the parameters $(E(\ket{\psi_s}),\alpha,-\beta,\gamma,\beta')$. Thus, the first two terms in Eq.\ (\ref{e6}) take the same value for both $|\psi\rangle$ and $|\psi'\rangle$, while the last one differs in sign. Hence, $E_{1|23}$ takes different values for these states and therefore $E_6$ distinguishes them, unless the last term in Eq.\ (\ref{e6}) is zero. We show in the following that this is the case iff $|\psi\rangle$ and $|\psi'\rangle$ are LU--equivalent. If $\beta,\beta'=0,\pi/2$ $|\psi\rangle$ and $|\psi'\rangle$ are clearly LU--equivalent. If $\alpha+\gamma=\pi/2$ or $\alpha-\gamma=\pi/2$ they are LU--equivalent, as can be seen by using that $Y(\beta)Z(\pi/2)=Z(\pi/2)Y(-\beta)$ together with the LU--equivalences of $\{\alpha,\beta,\gamma,\beta'\}$ given in Sec.\ I. It remains to consider the case $a=0$ or $b=0$. Then, $\ket{\psi}$ is LU-equivalent to $\sqrt{p}\ket{0}\ket{00}+\sqrt{1-p}\ket{1}U_2(\alpha,\beta,\gamma)\otimes Y(\beta')\ket{00}$, which is obviously LU--equivalent to $\ket{\psi'}$, since $\sigma_z\ket{0}=\ket{0}$ and $Y(\beta')\sigma_z=\sigma_zY(-\beta')$.


\subsection{Appendix D: Characterization of CLU and NCLU classes}

The aim of this section is to prove the characterization of the CLU class. We will first derive a mathematical characterization (see Lemma 1 below), which will then be used to classify CLU states according to the value the measure $E_1$ takes (Theorem 1 below).

As in the main text we consider without loss of generality the following decomposition of $\ket{\psi}$:
\bea \label{dec} \ket{\psi}=\sqrt{p}\ket{0}\ket{\psi_0}+\sqrt{1-p}\ket{1}\ket{\psi_1},\eea
where $\ket{\psi_0}=a_1\ket{00}+a_4\ket{11}$, with $a_1,a_4\in\mathbb{R}$ and $\ket{\psi_1}=b_1\ket{00}+b_2\ket{01}+b_3\ket{10}+b_4\ket{11}$, with $b_i\in  \mathbb{C}$. Moreover the following conditions are satisfied: \begin{enumerate} \item[C1)] $\langle \psi_i\ket{\psi_j}=\delta_{i,j}$. \item[C2)] $c_0,c_1\geq0$. 
\end{enumerate} Note that such a decomposition is always possible since a local phase gate, $\mbox{diag}(e^{i\alpha_0/2},e^{i\alpha_1/2})$ on the first qubit changes $c_0$ ($c_1$) to $c_0 e^{i\alpha_0}$ ($c_1 e^{i\alpha_1}$) respectively. Given this decomposition $\tilde{c}=-a_1b_4-a_4 b_1$. We are going to show now that $\tilde{c}$ is real or purely imaginary iff $\ket{\psi}$ is CLU.

\begin{lemma} For a state $\ket{\psi}$ there exists a decomposition as in Eq.\ (\ref{dec}) with $\tilde{c}\in\mathbb{R}$ or $\tilde{c}\in i\mathbb{R}$ iff $\ket{\psi}$ is CLU.
\end{lemma}
\textit{Proof.} First, notice that it suffices to prove that the state is CLU iff there exists a decomposition with $c_0,c_1,\tilde{c}\in\mathbb{R}$. This is because the latter condition holds iff there exists a decomposition with $c_0,c_1\geq0$ and $\tilde{c}\in\mathbb{R},i\mathbb{R}$ as a local phase transformation $c_k\rightarrow c_k e^{i\pi}$ ($k=0$ or $k=1$) induces $\tilde{c}\rightarrow\tilde{c}e^{i\pi/2}$. Also, we note that the condition that $0=\langle \psi_0\ket{\psi_1}=a_1b_1+a_4b_4$ implies that the phase of $b_1$ equals the one of $b_4$ (unless $c_0=0$, which will be studied separately). We write $b_1=b_1^r e^{i\phi}$ and $b_4=b_4^r e^{i\phi}$, where both, $b_1^r$ and $b_4^r$ are real. Let us start by proving the "only if" part. That is, we assume that $\tilde{c}=-a_1b_4-a_4b_1$ is real and show that this implies that $\ket{\psi}$ is CLU. $\tilde{c}$ being real implies that either $\phi=k \pi$ for some integer $k$ or $\tilde{c}=0$. In the first case, $c_1=2(-b^r_1b^r_4+b_2b_3)$, which is real iff $b_2b_3$ is real. Hence, $b_2=b_2^r e^{i\chi}$ and $b_3=b_3^r e^{-i\chi}$, where $b_{2,3}^r$ are real. It is then easy to see that the local unitary $\one\otimes Z(-\chi/2)\otimes  Z(\chi/2)$ transforms $\ket{\psi}$ into a state with real coefficients, which proves that the state is CLU. If $\tilde{c}=-a_1b_4-a_4b_1=0$ and $a_1^2\neq a_4^2$, C1 further implies that $b_1=b_4=0$ 
. Then, it can be easily shown again that the state can be mapped by local phase gates to a state with real coefficients, which is CLU. If $a_1^2= a_4^2$, then without loss of generality we can take $a_1=a_4=1/\sqrt{2}$, i.\ e.\ $|\psi_0\rangle=|\phi^+\rangle$. In this case C1 does not add any further constraint and we have $b_4=-b_1$, i.\ e.\ $\ket{\psi_1}=b_1\ket{00}+b_2\ket{01}+b_3\ket{10}-b_1\ket{11}$. We will show again that there exists local unitaries that transform $|\psi\rangle$ into a state with real coefficients.  Let $X=b_1|0\rangle\langle0|+b_2|0\rangle\langle1|+b_3|1\rangle\langle0|-b_1|1\rangle\langle1|$ and let $X=U\Sigma V^\dag$ be its singular value decomposition. This means that the local unitaries $U^\dag$ and $V^T$ transform $|\psi_1\rangle$ into its Schmidt form, i.\ e.\ $U^\dag\otimes V^T|\psi_1\rangle=|\psi_1^s\rangle\in\mathbb{R}^4$. Notice that the fact that $\tr X=0$ implies that $\tr(\Sigma V^\dag U)=0$ and, hence, $V^\dag U=e^{i\phi}W$, where $W$ is a unitary matrix with real diagonal entries $w_{11}$ and $w_{22}$ and off-diagonal entries of the form $w_{12}=w_{12}^re^{i\xi}$ and $w_{21}=w_{21}^re^{-i\xi}$ with $w_{12}^r$ and $w_{21}^r$ both real. Let $|\tilde{\psi}\rangle=\one\otimes Z(-\alpha)U^\dag\otimes Z(\alpha)V^T|\psi\rangle$. Using that the local unitary transformation $Z(\alpha)\otimes Z(-\alpha)$ for any $\alpha$ leaves any state in the Schmidt form invariant and that $U_1\otimes U_2|\phi^+\rangle=\one\otimes U_2U_1^T|\phi^+\rangle$ for all unitaries $U_{1,2}$, we have that
\begin{equation}
|\tilde{\psi}\rangle=\sqrt{p}e^{-i\phi}|0\rangle\one\otimes Z(\alpha)W^*Z(-\alpha)|\phi^+\rangle+\sqrt{1-p}|1\rangle|\psi_1^s\rangle.
\end{equation}
Taking into account the above conditions on $W$, we have that $Z(\alpha)W^*Z(-\alpha)$ is a real matrix if $\alpha$ is chosen such that $\alpha=\xi/2$. Therefore, the local unitary $diag(e^{i\phi},1)\otimes Z(-\xi/2)U^\dag\otimes Z(\xi/2)V^T$ transforms $|\psi\rangle$ into a state with real coefficients as we wanted to show.

It remains to consider the case $c_0=0$. We choose without loss of generality $a_4=0$. Then, condition (C1) implies that $b_1=0$ and therefore $\tilde{c}=-b_4$, which is real iff $b_4$ is. Condition C2 implies that $b_2b_3$ is real. In this case the local unitary $\one\otimes Z(-\chi/2)\otimes  Z(\chi/2)$ for some proper choice of $\chi$ transforms $\ket{\psi}$ into a state with real coefficients, which completes the proof of the "only if" part. 

Let us now show the implication in the opposite direction. If $\ket{\psi}$ is CLU, it has been shown in \cite{acin} that, then, there exists a product basis in which the state has real coefficients, i.\ e.\ $|\psi\rangle\simeq_{LU}\sum_{ijk} \lambda_{ijk} |ijk\rangle$ with $\la_{ijk}\in\mathbb{R}$ $\forall$ $i,j,k$. Thus, the singular value decomposition of the (real) matrix $X=\sum_{ijk}\lambda_{ijk}|i\rangle\langle jk| $ is
real and, therefore, so is the Schmidt decomposition of the state in the
splitting $1|23$. Hence, $c_0$, $c_1$ and $\tilde{c}$ can be chosen to be real.

\hfill$\square$

We are now in the position to prove that the CLU class can be characterized via the values the entanglement measure $E_1=E(|\psi_s\rangle)$ takes. Let us recall here the decomposition [Eq.\ (2) in the main text] we have proven in the main text, as we will be using it throughout this Supplemental Material. Any 3-qubit state can be written up to LUs as
\begin{equation}\label{sform}
|\psi\rangle=\frac{1}{\sqrt{2}}(|0\rangle_1|\psi_s\rangle_{23}+|1\rangle_1U_2\otimes U_3|\psi_s\rangle_{23}),
\end{equation}
where $|\psi_s\rangle$ has Schmidt decomposition, $U_2=Z(\alpha)Y(\beta)Z(\gamma)$, $U_3=Y(\beta')$, $E(\psi_s)\in [\mathfrak{E}(C_{23}),\mathfrak{E}(C^a_{23})]$, $\alpha,\gamma\in(-\pi/2,\pi/2]$ and $\beta,\beta'\in[0,\pi/2]$.

\begin{theorem}
$|\psi\rangle\in$ CLU iff there exists a decomposition of the form (\ref{sform}) such that either $E_1=\mathfrak{E}(C_{23})$ or $E_1=\mathfrak{E}(C_{23}^a)$.
\end{theorem}

\textit{Proof.} First, it should be noticed that, since $C(\rho_{23})=s_1-s_2$ \cite{concurrence} and $C^a(\rho_{23})=s_1+s_2$ \cite{ca} where $\{s_i\}$ are the singular values of $\tau$ (arranged as usual in non-increasing order), it can be easily seen that $C_{23}=C_-$ and $C_{23}^a=C_+$ with
\begin{equation}\label{cplusminus}
C_\pm^2=p^2c_0^2+(1-p)^2c_1^2+2p(1-p)(|\tilde{c}|^2\pm|c_0c_1-\tilde{c}^2|).
\end{equation}
On the other hand, we have that
\begin{equation}\label{ed}
C(\psi_s)^2=p^2c_0^2+(1-p)^2c_1^2+2p(1-p)(c_0c_1\cos2\omega+2|\tilde{c}|^2),
\end{equation}
with $\omega$ given by Eq.\ (3) in the main text. From this, it is then clear that if $\tilde{c}\in\mathbb{R}$ or $\tilde{c}\in i\mathbb{R}$, then either $C(\psi_s)=C_+$ or $C(\psi_s)=C_-$, which together with Lemma 1 proves the {\it only if} part of the theorem. To see that the implication in the opposite direction holds as well, we have to prove that $\arg\tilde{c}=n\pi,n\pi/2$ are the \textit{only} solutions to $C(\psi_s)=C_\pm$ under the constraint (3) of the main text. This constraint can be rewritten as
\begin{equation}
\cos2\omega=\frac{1-x^2-(1+x^2)\cos(2\arg\tilde{c})}{1+x^2-(1-x^2)\cos(2\arg\tilde{c})},
\end{equation}
where we use the notation $x=(pc_0+(1-p)c_1)/(pc_0-(1-p)c_1)$. Using this and writing $y=\cos(2\arg\tilde{c})$, $C(\psi_s)=C_\pm$ boils down to
\begin{align}
c_0c_1&\frac{1-x^2-(1+x^2)y}{1+x^2-(1-x^2)y}+|\tilde{c}|^2\nonumber\\
&=\pm\sqrt{c_0^2c_1^2+|\tilde{c}|^4-2c_0c_1|\tilde{c}|^2y}.
\end{align}
After squaring this last expression, our problem reduces to identifying the zeros of a third order polynomial $p(y)$. More precisely, we know two zeros $y_1=-1$ and $y_2=1$ (which correspond to $\arg\tilde{c}=n\pi,n\pi/2$) and we have to check that the third one $y_3$ fulfills $y_3\notin(-1,1)$, so that $y_3=\cos(2\arg\tilde{c})$ cannot be inverted and, hence, there are not more solutions to $C(\psi_s)=C_\pm$. Indeed, using that $|p(0)|=|\gamma y_1y_2y_3|$ where $\gamma$ is the leading coefficient of $p$, we see that
\begin{equation}
|y_3|=\frac{|2c_0c_1x^2+|\tilde{c}|^2(x^4-1)|}{|\tilde{c}|^2(x^2-1)^2}.
\end{equation}
Now, since we take $c_0,c_1\geq0$ (which also implies that $x\geq1$), it is easily seen that $|y_3|\geq1$, which completes the proof.

\hfill$\square$

\subsection{Appendix E: Characterization of the 4 subclasses inside the CLU class}

In this section we will prove the characterization of the four subclasses in which the CLU class can be divided according to the particular value of $E_1$ (which by Theorem 1 above can be either $\mathfrak{E}(C_{23})$ or $\mathfrak{E}(C_{23}^a)$).

The fact that class 1 ($E_1=\mathfrak{E}(C_{23})=\mathfrak{E}(C_{23}^a)$) corresponds to the W class can be easily seen as follows. It can be shown that $(C_{23}^a)^2=C_{23}^2+\tau_{123}$ (see e.\ g.\ \cite{chi}), where $\tau_{123}$ is the tangle. Since the W class is characterized by a vanishing tangle, the result follows straightforwardly.

The identification of the other three classes will require more work. As Lemma $2$ below shows, the sign of $\det\tau$ will play a crucial role in this task. We recall that the matrix $\tau$ associated with a state $|\psi\rangle$ is a function of its reduced state $\rho_{23}$. It is given by $\tau_{ij}=\sqrt{p_ip_j}\langle\tilde{\psi_i}|\psi_j\rangle$, where $\{p_i\}$ are the eigenvalues of $\rho_{23}$ and $\{|\psi_i\rangle\}$ the corresponding eigenstates. Therefore, it is symmetric. Since $\det\tau=0$ iff the tangle $\tau_{123}=0$ (as follows from \cite{ckw}), which corresponds to class 1, we assume in the following that $\det\tau\neq 0$. First note, that by redefining the sign of $c_i\in\mathbb{R}$ we can always obtain a decomposition with $\tilde{c}\in\mathbb{R}$ (recall that we are dealing with the CLU class and, hence, Lemma 1 applies). Thus, $\tau\in\mathbb{R}^{2\times2}$ in what follows unless otherwise stated.

Before relating the sign of $\det\tau$ to our classes, we need to discuss when this property is well defined. This is because it could in principle happen that two LU equivalent states lead to different signs for $\det\tau$. We now prove that this is indeed the case iff there is a choice of eigenstates $|\psi_0\rangle$ and $|\psi_1\rangle$ such that $\tilde{c}=0$. First, if the latter condition holds, then it is clear that under a different choice of global phase for $|\psi_0\rangle$, $|\psi_0\rangle\rightarrow i|\psi_0\rangle$, $\tau$ remains real and the sign of its determinant is flipped. To prove the implication in the other direction we have to consider how $\tau$ transforms under the LU transformations $|\psi\rangle\rightarrow W_1\otimes W_2\otimes W_3|\psi\rangle$. Using that $\tau=M_\psi^\ast \sigma_y\otimes \sigma_y M_\psi^\dagger$, where $M_\psi=\sqrt{p}\ket{0}\bra{\psi_0}+\sqrt{1-p}\ket{1}\bra{\psi_1}$, and that $W \sigma_y W^T=\det (W)\sigma_y$ for any matrix $W$, we find that $\tau$ transforms to
\begin{equation}\label{tautilde}
\tilde{\tau}=\det(W_2)\det(W_3)W_1\tau W_1^T.
\end{equation}
Now, we have to distinguish the following two cases. i) $p\neq1/2$ and, hence, the eigenspace of $\rho_{23}$ is nondegenerate and the Schmidt form unique. Then, $W_1$ can only take the form $W_1=diag(e^{i\alpha_1},e^{i\alpha_2})$ to preserve the orthogonality of $|\psi_0\rangle$ and $|\psi_1\rangle$. In this case $\det\tilde{\tau}=-\det\tau$ iff $\det(W_2)\det(W_3)e^{i(\alpha_1+\alpha_2)}=\pm i$. However, notice that for the off-diagonal entry we have that $\tilde{\tau}_{12}=\det(W_2)\det(W_3)e^{i(\alpha_1+\alpha_2)}\tau_{12}$ and, hence, the fact that both $\tau$ and $\tilde{\tau}$ should be real implies that $\tilde{c}=0$ has to hold. ii) $p=1/2$. Then, the eigenspace of $\rho_{23}$ is degenerate and the Schmidt form not unique (as $|\psi_0\rangle$ and $|\psi_1\rangle$ can be chosen to be any orthonormal basis of the eigenspace). Thus, $W_1$ might be an arbitrary unitary matrix. We show now that in this case there always exist LUs $W_1\otimes W_2\otimes W_3$ such that the sign of $\det\tau$ can be changed and such that $\tilde{c}=0$. Let $V$ denote the real orthogonal matrix that diagonalizes $\tau$ and let us choose $W_1=\left(
                                                                              \begin{array}{cc}
                                                                                i & 0 \\
                                                                                0 & 1 \\
                                                                              \end{array}
                                                                            \right)V$ and $W_2=W_3=\one$ in Eq.\ (\ref{tautilde}). The matrix $W_1$ defined in this way is such that $\det W_1=\pm i$ (thus $\det\tilde{\tau}=-\det\tau$) and with this choice $\tilde{\tau}$ is real and, furthermore, diagonal. Hence, there exists a choice for which $\tilde{c}=0$. This ends the proof.

Therefore, special care must be taken with those states that allow for $\tilde{c}=0$. This is not surprising, as we have seen in Sec.\ I that these are the only ones for which the choice of $|\psi_s\rangle$ in Eq.\ (\ref{sform}) is in principle not unique. 
Let us now show how the sign of $\det\tau$ is related to the different classes. 

\begin{lemma} There exists a decomposition of the form (\ref{sform}) such that $E_1=\mathfrak{E}(C_{23})$ iff $\det\tau\geq0$ and such that $E_1=\mathfrak{E}(C_{23}^a)$ iff $\det\tau\leq0$.
\end{lemma}
\textit{Proof.} Since $\tau\in\mathbb{R}^{2\times2}$, the unitary in Eq.\ (1) in the main text that allows for the transformation to the decomposition (\ref{sform}) can be taken to be $U=\left(\begin{array}{cc}                                                                                                                                                            1 & i \\
1 & -i \\                                                                                                                                                             \end{array}
\right)/\sqrt{2}$. We will also use that $|\det\tau|=\tau_{123}/4$, which follows from the results of \cite{ckw}. With this in mind,
\begin{equation}\label{tauu}
\tau_{123}/4=|\det\tau|=|\det(U\tau U^T)|=\frac{1}{4}|C_1^2-(\tr\tau)^2|,
\end{equation}
where $C_1$ is such that $E_1=\mathfrak{E}(C_1)$.
On the other hand, since $\tau$ is symmetric (and now real), it is Hermitian and, hence, the absolute value of its eigenvalues equal the singular values. Therefore, either $|\tr\tau|=s_1-s_2=C_{23}$ (when $\det\tau\leq0$) or $|\tr\tau|=s_1+s_2=C_{23}^a$ (when $\det\tau\geq0$). Now, since $(C_{23}^a)^2=C_{23}^2+\tau_{123}$ has to hold in Eq.\ (\ref{tauu}), the result follows. (This can also be seen by comparing Eqs.\ (\ref{cplusminus}) and (\ref{ed})).
\hfill$\square$

To identify the sign of $\det\tau$ we will make intensive use of the standard form (up to LUs) for 3-qubit states presented in \cite{acin},
\begin{equation}\label{toni}
|\psi\rangle=\lambda_0|000\rangle+\lambda_1e^{i\phi}|100\rangle+\lambda_2|101\rangle+\lambda_3|110\rangle+\lambda_4|111\rangle,
\end{equation}
where $\lambda_i\geq0$ $\forall i$ and $\phi\in[0,\pi]$. Some equations might be more compactly written in terms of the polynomial invariants \cite{acin}
\begin{align}
J_1&=|\la_1\la_4e^{i\phi}-\la_2\la_3|^2,\nonumber\\
J_2&=\la_0^2\la_2^2,\nonumber\\
J_3&=\la_0^2\la_3^2,\nonumber\\
J_4&=\la_0^2\la_4^2,\nonumber\\
J_5&=\la_0^2(J_1+\la_2^2\la_3^2-\la_1^2\la_4^2). \label{polinv}
\end{align}
Grassl's invariant, which discriminates $|\psi\rangle$ and $|\psi^*\rangle$ for states in the NCLU class, is given by
\begin{equation}\label{grassl}
J_6=\la_0^4\la_4^2(\la_4(1-2\la_0^2-2\la_1^2)+2\la_1\la_2\la_3e^{-i\phi})^2.
\end{equation}
According to \cite{acin}, the CLU class can be characterized by certain conditions satisfied by the polynomial invariants [Eq.\ (\ref{polinv})]. Namely, a state is in the CLU class iff either
\begin{equation}\label{CLU1}
|J_5|=2\sqrt{J_1J_2J_3}
\end{equation}
or
\begin{equation}\label{CLU2}
(J_4+J_5)^2-4(J_1+J_4)(J_2+J_4)(J_3+J_4)=0.
\end{equation}
The states satisfying Eq.\ (\ref{CLU1}) are moreover shown to be equivalent to those for which the standard form given in Eq.\ (\ref{toni}) is already real (i.\ e.\ $\phi=0,\pi$) \cite{acin}. Alternatively, these two subclasses can be characterized by the value of $J_6$. Since they are CLU, $J_6(\psi)=J_6(\psi^*)$ and it follows that $J_6$ must be real. The subclass characterized by Eq.\ (\ref{CLU1}) corresponds to $J_6\geq0$ (the expression which is squared in Eq.\ (\ref{grassl}) is real) and the subclass characterized by Eq.\ (\ref{CLU2}) to $J_6\leq0$ (the expression which is squared in Eq.\ (\ref{grassl}) has to be purely imaginary). Furthermore, it is shown in \cite{acin} that the states for which Eq.\ (\ref{CLU1}) holds can be written up to LUs as
\begin{equation}\label{class2}
c_1|000\rangle+c_2|\phi_1 \phi_2 \phi_3\rangle
\end{equation}
with $ c_i\in\mathbb{R}$ and $\phi_i \in \mathbb{R}^2$, which correspond to our class 2. On the other hand it is also shown therein that the states satisfying Eq.\ (\ref{CLU2}) are LU equivalent to states of the form
\begin{equation}\label{class3}
|000\rangle+e^{i\delta}|\phi_1 \phi_2 \phi_3\rangle
\end{equation}
up to normalization, where $\delta\in \mathbb{R}$ and $\phi_i \in \mathbb{R}^2$. Notice that this class corresponds to our class 3 \footnote{Here we ignore the W class which fulfills both Eqs.\ (\ref{CLU1}) and (\ref{CLU2}) which, as shown above, is characterized by $\det\tau=0$.}.

Our aim is to connect these families with the sign of $\det\tau$. More precisely, we will show in the following that for states for which Eq.\ (\ref{CLU1}) holds, i.\ e.\ they can be written as in Eq.\ (\ref{class2}), (Eq.\ (\ref{CLU2}) holds, i.\ e.\ they can be written as in Eq.\ (\ref{class3})) then $\det\tau\leq0$ ($\det\tau\geq0$). Using then Lemma 2, we have that any state in class 2 (class 3) fulfills $E_1=\mathfrak{E}(C_{23}^a)$ ($E_1=\mathfrak{E}(C_{23})$). Since conditions (\ref{CLU1})--(\ref{CLU2}) characterize completely the CLU class, the implication has to hold as well in the opposite direction, i.\ e.\ if $E_1=\mathfrak{E}(C_{23}^a)$ ($E_1=\mathfrak{E}(C_{23})$) then the state can be written as in Eq.\ (\ref{class2}) (Eq.\ (\ref{class3})). Thus, this will prove the characterization of classes 2 and 3. Of course, this makes sense only if the sign of $\det\tau$ is well defined. Hence, for this task we will assume that a choice of $\tau$ for which $\tilde{c}=0$ does not exist. We will see afterwards that this assumption is justified as we will prove that $\tilde{c}$ can be taken to be zero iff both conditions (\ref{CLU1}) and (\ref{CLU2}) hold \textit{at the same time}. This will finally lead to the characterization of class 4 which is given by the intersection of classes 2 and 3.

First, we need to write $\tau$ in terms of the $\la$'s. Hence, we need to find the $1|23$ Schmidt decomposition of the state given in Eq.\ (\ref{toni}). Notice that Lemma 2 holds when the global phases of the Schmidt eigenstates for $23$, $|\psi_0\rangle$ and $|\psi_1\rangle$, are chosen such that $\tau\in\mathbb{R}^{2\times2}$ (which is always possible for CLU states by Lemma 1). Thus, once $\tau$ is computed from the Schmidt decomposition of the state of Eq.\ (\ref{toni}) a further transformation might be needed for the latter condition to hold. As before, we have that these global phase transformations transform the entries of $\tau$ as
\begin{equation}\label{phasetrans}
c_0\rightarrow c_0e^{i\alpha},\quad c_1\rightarrow c_1e^{i\beta},\quad\tilde{c}\rightarrow \tilde{c}e^{i(\alpha+\beta)/2}.
\end{equation}
We will denote by $\tau_r$ the real matrix obtained by these transformations. However, fortunately, since, as mentioned above, for the subclass characterized by Eq.\ (\ref{CLU1}) it holds that $\phi=0,\pi$, we will always deal with real expressions in this case. Hence, we will only need to take care of this fact for the second subclass of states satisfying Eq.\ (\ref{CLU2}).

Let us start by considering the case in which Eq.\ (\ref{CLU1}) holds and Eq.\ (\ref{CLU2}) does not. We have to show that this implies that $\det\tau_r\leq0$. First, we consider that $\la_1=0$ (which agrees with the fact that Eq.\ (\ref{CLU1}) is satisfied), as in this case the state in Eq.\ (\ref{toni}) has already the Schmidt form. Thus, $p=\la_0^2$, $c_0=0$, $c_1=2\la_2\la_3/(1-p)$ and $\tilde{c}=-\la_4/\sqrt{1-p}$. Thus, $\tau$ is then already real and $\det\tau_r=-J_4\leq0$ as we wanted to show. Notice that here one should exclude $\la_0^2=1/2$, as this would imply using Eqs.\ (\ref{polinv}) that Eq.\ (\ref{CLU2}) would hold as well. This is not surprising since this corresponds to $p=1/2$ for which we have seen that the sign of $\det{\tau_r}$ can be changed by LU transformations. We will see shortly that the $\la_1=0$ case is the only possibility in which $p=1/2$ may hold.

We now consider the case when $\la_1\neq0$. To compute the Schmidt form, we need to find the singular value decomposition $X=U\Sigma V^\dag$ of
\begin{equation}\label{svd}
X=\left(
    \begin{array}{cccc}
      \la_0 & 0 & 0 & 0\\
      \la_1e^{i\phi} & \la_2 & \la_3 & \la_4 \\
    \end{array}
  \right),
\end{equation}
$U$ providing the Schmidt basis for qubit 1, the singular values $\sigma_\pm$ providing the Schmidt coefficients (i.\ e.\ $\sigma_+=p, \sigma_-=1-p$) and the first two columns of V providing the Schmidt basis for qubits 2 and 3 (i.\ e.\ $v_1=|\psi_0\rangle$ and $v_2=|\psi_1\rangle$). After some algebra one finds that
\begin{align}
\sigma_\pm&=\frac{1\pm\sqrt{1-4(J_2+J_3+J_4)}}{2},\nonumber\\
U&=\left(
    \begin{array}{cc}
      -\la_0\la_1e^{i\phi}/k_+ & -\la_0\la_1e^{i\phi}/k_- \\
      (\la_0^2-\sigma_+)/k_+ & (\la_0^2-\sigma_-)/k_- \\
    \end{array}
  \right),
\end{align}
where $k_\pm$ are just normalization factors to make the columns of $U$ of unit norm (i.\ e.\ $k_\pm^2=\la_0^2\la_1^2+(\la_0^2-\sigma_\pm)^2$). Notice that the case $\sigma_{\pm}=1/2$ does not occur here as $J_2+J_3+J_4=1/4$ implies, using Eqs.\ (\ref{polinv}), that $\lambda_0^2(1-\lambda_1^2-\lambda_0^2)=1/4$, and therefore that $\lambda_1=0$. As $V\Sigma^\dag=X^\dag U$, we then have that
\begin{equation}
\sqrt{\sigma_\pm}v_\pm=\left(
  \begin{array}{c}
    -\la_0^2\la_1e^{i\phi}/k_\pm+(\la_0^2-\sigma_\pm)\la_1e^{i\phi}/k_\pm \\
    \la_2(\la_0^2-\sigma_\pm)/k_\pm \\
    \la_3(\la_0^2-\sigma_\pm)/k_\pm \\
    \la_4(\la_0^2-\sigma_\pm)/k_\pm \\
  \end{array}
\right),
\end{equation}
where $v_1=v_+$ and $v_2=v_-$. From this, it follows that
\begin{align}
k_\pm\sigma_\pm c_{0,1}&=2\la_0^2\la_1\la_4(\la_0^2-\sigma_\pm)e^{i\phi}\nonumber\\
&-2(\la_0^2-\sigma_\pm)^2(\la_1\la_4e^{i\phi}-\la_2\la_3),\nonumber\\
k_+k_-\sqrt{\sigma_+\sigma_-}\,\tilde{c}&=\la_0^2\la_1\la_4(2\la_0^2-1)e^{i\phi}\nonumber\\
&+2\la_0^2\la_1^2(\la_1\la_4e^{i\phi}-\la_2\la_3),\label{algo2}
\end{align}
and, moreover,
\begin{align}
k_+^2k_-^2\sigma_+\sigma_-c_0c_1&=-4\la_0^6\la_1^4\la_4^2e^{2i\phi}+A,\nonumber\\\label{algo}
k_+^2k_-^2\sigma_+\sigma_-\tilde{c}^2&=\la_0^4\la_1^2\la_4^2(4\la_0^4-4\la_0^2+1)e^{2i\phi}+A,
\end{align}
where
\begin{align}
A&=4\la_0^4\la_1^4\la_4^2(2\la_0^2+\la_1^2-1)e^{2i\phi}\nonumber\\
&+4\la_0^4\la_1^3\la_2\la_3[\la_4(1-2\la_0^2-2\la_1^2)e^{i\phi}+\la_1\la_2\la_3].
\end{align}
One then finally finds that
\begin{equation}\label{dettau}
k_+^2k_-^2\det\tau=4\la_0^4\la_1^2\la_4^2(J_2+J_3+J_4-\frac{1}{4})e^{2i\phi}.
\end{equation}

As discussed above, then for the states for which Eq.\ (\ref{CLU1}) is true it holds that $k_+^2k_-^2\det\tau_r=4\la_0^4\la_1^2\la_4^2(J_2+J_3+J_4-1/4)$. Since it can be shown that $J_2+J_3+J_4\leq1/4$ \cite{acin}, we have that $\det\tau_r\leq0$ in this case. This finishes the identification of class 2.

Let us now consider the case for which Eq.\ (\ref{CLU1}) does not hold (i.\ e.\ $\phi\neq0,\pi$) but for which Eq.\ (\ref{CLU2}) does \footnote{Notice that it suffices to assume that Eq.\ (\ref{CLU1}) does not hold since, given that we are considering CLU states, this automatically implies that Eq.\ (\ref{CLU2}) must be fulfilled.}. Hence, we need to prove that $\det\tau_r\geq0$. Notice that $\la_1=0$ implies that Eq.\ (\ref{CLU1}) is true. Therefore, we just need to consider the $\la_1\neq0$ case which, as above, leads us to Eq.\ (\ref{dettau}). However, as discussed before, to construct $\tau_r$ for this subclass, a specific transformation of the form (\ref{phasetrans}) must be performed. That is, we choose the global phase of $\ket{\psi_0}$ ($\alpha/2$) and the global phase of $\ket{\psi_1}$ ($\beta/2$) such that $\tau$ gets real. Since under such a transformation $\det\tau$ changes to $\det\tau e^{i(\alpha+\beta)}$, we get a real expression in (\ref{dettau}) iff either $\alpha+\beta=-2\phi$ (which would lead to $\det\tau\leq0$) or $\alpha+\beta=\pi-2\phi$ (which would lead to $\det\tau\geq0$). Moreover, $\tilde{c}e^{i(\alpha+\beta)/2}$ must be real, i.e. $\tilde{c}^2e^{i(\alpha+\beta)}\geq0$ must hold. In other words, the sign of $\tilde{c}^2e^{-2i\phi}$ will tell us whether $\alpha$ and $\beta$ have to be chosen such that $\alpha+\beta=-2\phi$ or $\alpha+\beta=\pi-2\phi$. In the following we prove that for any choice of the phases $\alpha$, $\beta$ which lead to $\tau\in\mathbb{R}^{2\times2}$, it holds that $\tilde{c}^2e^{-2i\phi}\leq0$. This means that $\alpha+\beta=\pi-2\phi$ is the proper choice to obtain $\tau_r$. Hence, for the CLU states such that $\phi\neq0,\pi$ (i.\ e.\ those satisfying Eq.\ (\ref{CLU2})), it holds that $\det\tau_r\geq0$ and therefore these states belong to class $3$. To see that $\tilde{c}^2e^{-2i\phi}\leq0$, notice that the transformation (\ref{phasetrans}) must fulfill both $c_0c_1\in\mathbb{R}$ and $\tilde{c}^2\in\mathbb{R}$ \footnote{Notice that $c_0c_1\in\mathbb{R}$ already implies that both $c_0$ and $c_1$ can be taken to be real. This is because this means that $c_0=c_0^re^{i\varphi}$ and $c_1=c_1^re^{-i\varphi}$ where $c_0^r$ and $c_1^r$ are both real. Hence, a further global phase transformation of the form (\ref{phasetrans}) can be taken, which makes $c_0,c_1\in\mathbb{R}$ without changing $\tilde{c}$.}. Equation (\ref{algo}) implies that for the states for which this is possible $Ae^{-2i\phi}\in\mathbb{R}$ must hold. This means that these states are such that
\begin{equation}
\label{Eq35}
\la_4(1-2\la_0^2-2\la_1^2)\sin\phi+\la_1\la_2\la_3\sin2\phi=0.
\end{equation}
Since now $\phi\neq0,\pi$, Eq.\ (\ref{Eq35}) is equivalent to
\begin{equation}\label{connectioni6}
\la_4(1-2\la_0^2-2\la_1^2)+2\la_1\la_2\la_3\cos\phi=0
\end{equation}
or, simply \footnote{Comparing with Eq. (\ref{grassl}), Eq.\ (\ref{connectioni6}) shows that these states indeed correspond to the subclass given by Eq.\ (\ref{CLU2}) since $J_6\leq0$.}
\begin{equation}\label{cos}
\cos\phi=-\frac{\la_4(1-2\la_0^2-2\la_1^2)}{2\la_1\la_2\la_3}.
\end{equation}
Inserting Eq. (\ref{Eq35}) in Eq. (\ref{algo}) leads to
\begin{align} \label{Eq36}
&k_+^2k_-^2\sigma_+\sigma_-\tilde{c}^2e^{-2i\phi}=4\la_0^4\la_1^2\la_4^2[1/4-(\la_0^2+\la_1^2)(1-\la_0^2-\la_1^2)]\nonumber\\
&+4\la_0^4\la_1^3\la_2\la_3[\la_4(1-2\la_0^2-2\la_1^2)\cos\phi+\la_1\la_2\la_3\cos2\phi].
\end{align}
 The second line of the above equation can be rewritten using Eq. (\ref{connectioni6}), which leads to
\begin{equation}
\la_4(1-2\la_0^2-2\la_1^2)\cos\phi+\la_1\la_2\la_3\cos2\phi=-\la_1\la_2\la_3.
\end{equation}
Using now that $
1/4-(\la_0^2+\la_1^2)(1-\la_0^2-\la_1^2)\leq(1-2\la_0^2-2\la_1^2)^2$ for arbitrary $\la_0, \la_1$ such that $\la_0^2+\la_1^2\leq 1$ we find the following upper bound for the left hand side of Eq.\ (\ref{Eq36})
\begin{align}
&k_+^2k_-^2\sigma_+\sigma_-\tilde{c}^2e^{-2i\phi}\nonumber\\\label{ineqs}
&\leq\la_0^4\la_1^2[\la_4^2(1-2\la_0^2-2\la_1^2)^2-4\la_1^2\la_2^2\la_3^2]\leq0,
\end{align}
where in the second inequality we have used Eq.\ (\ref{cos}). Thus, $\tilde{c}^2e^{-2i\phi}\leq 0$, which concludes the proof that for any state which does not fulfill Eq.\ (\ref{CLU1}) but does fulfill Eq.\ (\ref{CLU2}) $\det\tau\geq 0$ holds.

Finally, let us consider the case for which both Eqs.\ (\ref{CLU1}) and (\ref{CLU2}) hold, i.\ e.\ $J_6=0$. Recalling that we are excluding the class of W states already characterized as class 1 (for which these conditions are also true), this means that the states are LU equivalent to
\begin{equation}
|000\rangle+|\phi_1 \phi_2 \phi_3\rangle
\end{equation}
up to normalization and with $\phi_i \in \mathbb{R}^2$. This can be understood from the fact that no state in class 2 can be transformed to any in class 3 via LOCC. In particular, it cannot be LU equivalent to any state in class 3. Therefore, this class must contain the intersection of the forms given in Eqs.\ (\ref{class2}) and (\ref{class3}), which corresponds to our class 4. It remains to show that this class is the only one for which $\det\tau_r$ can take both signs. That is, leaving aside the W class, class 4 is characterized by the possibility of having $\tilde{c}=0$. We will prove that both Eqs.\ (\ref{CLU1}) and (\ref{CLU2}) hold iff there exists a choice of $\tau_r$ for which $\tilde{c}=0$. The proof of the \textit{if} part is immediate as can be seen as follows. If $p=1/2$ we have seen before that this already implies that $\tilde{c}=0$ is possible. If $p\neq1/2$ ,we have just seen that if Eq.\ (\ref{CLU1}) holds ($\phi=0,\pi$) the matrix $\tau$ constructed from the standard form (\ref{toni}) is already real and that $\tilde{c}^2\geq0$. If Eq.\ (\ref{CLU2}) is moreover satisfied, then it also holds that $\tilde{c}^2\leq0$ and, hence, $\tilde{c}=0$. To prove the implication in the other direction we assume that $\tilde{c}=0$. If $\la_1\neq0$, we can use Eq.\ (\ref{algo2}), which then yields that
\begin{equation}
\la_0^2[\la_4(1-2\la_0^2-2\la_1^2)+2\la_1\la_2\la_3e^{-i\phi}]=0.
\end{equation}
This means that $J_6=0$ and, thus, both Eqs.\ (\ref{CLU1})-(\ref{CLU2}) hold. Following our discussion before Eq.\ (\ref{svd}), in the case $\la_1=0$, there exists a choice of $\tau_r$ with $\tilde{c}=0$ iff either one of the two possibilities is fulfilled: $\la_4=0$ or $p=1/2$, for which the Schmidt form is not unique. The first one clearly leads to $J_6=0$ (and, in particular, to the W class since $J_4=\tau_{123}/4$ \cite{acin}). In the second case, $\la_0^2=1/2$ must hold as well and, therefore, $J_6=0$. This completes the proof.

Thus, as we discussed in Sec.\ I, since $\tilde{c}=0$, for the states in class 4 there exist in principle different choices for $|\psi_s\rangle$ in the decomposition (\ref{sform}).

\end{document}